\documentclass[twocol]{ametsoc}

\usepackage[]{verbatim} 	
\usepackage{tabularx}
\usepackage{graphicx}
\usepackage{wrapfig}
\usepackage{xcolor}
\usepackage{enumitem}
\usepackage{subcaption}
\usepackage{colortbl}
\usepackage{amsmath}

\journal{jas}

\bibpunct{(}{)}{;}{a}{}{,}

\title{An idealized physical model for the severe convective storm environmental sounding \\
{\color{red}Accepted in Journal of the Atmospheric Sciences 2020-10-26, there may still be copy-editing errors}}

    \authors{Daniel R. Chavas\correspondingauthor{Daniel R. Chavas, 
     Purdue University, Department of Earth, Atmospheric, and Planetary Sciences,
     550 Stadium Mall Drive HAMP 3221,
     West Lafayette, IN 47907.}}

     \affiliation{Purdue University, Department of Earth, Atmospheric, and Planetary Sciences, West Lafayette, IN}

\email{drchavas@gmail.com}

    \extraauthor{Daniel T. Dawson II}
    \extraaffil{Purdue University, Department of Earth, Atmospheric, and Planetary Sciences, West Lafayette, IN}

\abstract{This work develops a theoretical model for steady thermodynamic and kinematic profiles for severe convective storm environments, building off of the two-layer static energy framework developed in Agard and Emanuel (2017). The model is phrased in terms of static energy, and it allows for independent variation of the boundary layer and free troposphere separated by a capping inversion. An algorithm is presented to apply the model to generate a sounding for numerical simulations of severe convective storms, and the model is compared and contrasted with that of Weisman and Klemp. The model is then fit to a case-study sounding associated with the 3 May 1999 tornado outbreak, and its potential utility is demonstrated via idealized numerical simulation experiments. A long-lived supercell is successfully simulated with the historical sounding but not the analogous theoretical sounding. Two types of example experiments are then performed that do simulate a long-lived supercell: 1) a semi-theoretical experiment in which a portion of the theoretical sounding is modified to match the real sounding (low-level moisture); 2) a fully-theoretical experiment in which a model physical parameter is modified (free-tropospheric relative humidity). Overall, the construction of this minimal model is flexible and amenable to additional modifications as needed. The model offers a novel framework that may be useful for testing how severe convective storms depend on the vertical structure of the hydrostatic environment, as well as for linking variability in these environments to the physical processes that produce them within the climate system.}

\begin{document}

\maketitle

%

\section{Introduction}

While substantial advances have been made in the understanding and prediction of severe convective storms (SCS), operational predictability remains limited and thus substantial risks to life and property persist. From 1995-2015, tornadoes caused 1710 deaths in the U.S., ranking as the third-deadliest weather phenomenon (NOAA). Our ability to predict these weather risks in the current or future climate depends crucially on a physical understanding of the dependence of SCS events on their larger-scale environment. Forecasting and research applications have largely focused on bulk (i.e. vertically-integrated) thermodynamic and kinematic parameters as part of the successful ``ingredients-based'' framework for SCS environment diagnosis and forecasting \citep{Doswell_Brooks_Maddox_1996,Doswell_2001,Tippett_etal_2015}.

A principal focus of SCS research is supercells, which produce the majority of SCS-related hazardous weather, particularly significant tornadoes \citep{Duda2013}. Past work has demonstrated that supercells are associated with large magnitudes of CAPE and 0-6km bulk vertical wind shear \citep{Weisman_Klemp_1982,Weisman_Klemp_1984,Tippett_etal_2015}, with the latter being a better discriminator between supercell and non-supercell environments than the former \citep{Thompson2003,Thompson2007}. Additionally, the strength of the low-level storm-relative flow, which is correlated with 0-6km bulk shear magnitude \citep{Warren2017}, has been identified as a discriminator between supercell and non-supercell environments \citep{DroegemeierLazarusDaviesJones1993,PetersNowotarskiMorrison2019,Petersetal2020,PetersNowotarskiMullendore2020}. Hence, the product of CAPE and 0-6km bulk shear is commonly used as an environmental proxy for potential SCS activity \citep{Brooks_Lee_Craven_2003,Gensini_Ashley_2011,Seeley_Romps_2015}. Significant tornado events are further linked to high magnitudes of low-level storm-relative environmental helicity (SRH) and low values of the lifting condensation level (LCL) \citep{Brooks1994,Rasmussen1998,Thompson2003,Thompson_Edwards_Mead_2004}, which are combined with CAPE and 0-6 km shear in the Significant Tornado Parameter (STP) for the forecasting of strong tornadoes \citep{Thompson_Edwards_Mead_2004}. This ingredients-based approach using bulk parameters can also provide meaningful insight into the spatial and temporal distribution of SCS activity \citep{Gensini_Ashley_2011,Rasmussen_Houze_2016,Li_etal_2020}, including long-term spatial shifts in tornado activity \citep{Agee_etal_2016, Gensini_Brooks_2018}. Moreover, these bulk parameter proxies have been used to estimate changes in severe weather and tornado risk under future climate change \citep{Trapp_etal_2007,Trapp_Diffenbaugh_Gluhovsky_2009, Diffenbaugh_Scherer_Trapp_2013, Seeley_Romps_2015}. 


Nevertheless, details of the vertical thermodynamic and shear profiles not captured by bulk parameters are likely to play important roles in storm evolution. Which details within a particular sounding actually matter for the evolution of a severe convective storm?  The lack of understanding of the effects of such higher-order variability is likely an important contributor to reduced SCS predictability on daily and sub-daily time scales \citep[e.g, ][]{Elmore2002,Elmore2002a,Cintineo2013}. Moreover, bulk proxy statistical relationships trained on canonical high CAPE and high bulk shear environments may be inappropriately applied to non-canonical environments, such as ones with high bulk shear yet relatively low CAPE in which quasi-linear convective systems are common \citep{Sherburn2014a}. Finally, because bulk proxies are necessarily validated only against a relatively short historical record, their application to future climates is not only uncertain but potentially misleading if the chosen proxies do not correctly scale with actual SCS risk across climate states \citep{Trapp_etal_2011,Gensini_Mote_2015,Hoogewind_Baldwin_Trapp_2017,Trapp_Hoogewind_2016,Trapp_Hoogewind_LasherTrapp_2019}. The above issues indicate the need for a deeper physical understanding of the role of the vertical thermodynamic and kinematic structure for fixed values of a given bulk proxy. 


Because these bulk parameters are by definition vertically integrated measures of the environment, two environments can yield the same bulk value despite having very different vertical thermodynamic and kinematic structures \citep{McCaul_Weisman_2001,Petersetal2020}. \citet{Weisman_Klemp_1984} were among the first to investigate how SCS morphology and evolution depend on vertical environmental structure using a cloud-resolving numerical model (CRM). Central to their methodology was a parametric model of the vertical thermodynamic profile \citep[][hereafter WK]{Weisman_Klemp_1982}. Since then, many idealized CRM studies have used the WK profile to investigate different aspects of SCS and their environments. These include three categories of experiments: 1) parameter sweep studies varying CAPE and shear \citep{Kirkpatrick2011,Lawson_2019}; 2) the vertical distribution of buoyancy or shear at fixed values of CAPE and bulk shear, respectively \citep{McCaul_Weisman_2001,Kirkpatrick2009,Guarriello2018,Brown2019}; and 3) variations in parameters independent of bulk parameters, particularly free-tropospheric moisture \citep{Gilmore1998,James2006,James2010,McCaul2004,Honda2015}. From these and related studies, an improved understanding of how higher-order vertical variability of SCS environmental profiles is slowly emerging. This seminal work using idealized sounding models to test sensitivities of convective evolution represents the foundation that we build off of in this study. 

\begin{figure*}[t]
\centerline{\includegraphics[width=0.9\textwidth]{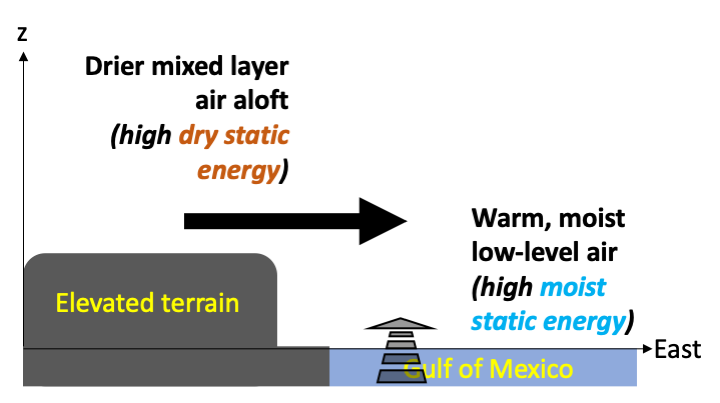}
\vspace{-4mm}}
\caption{Conceptual diagram of how an environment with large CAPE is generated east of the Rocky Mountains, in a static energy framework following Agard and Emanuel (2017).}

\label{fig:conceptual}
\vspace{-4mm}
\end{figure*}

While the WK thermodynamic sounding has been undeniably useful in advancing our understanding of basic storm dynamics over the past few decades, its construction is somewhat ad hoc -- its structure is composed of simple parametric equations for the tropospheric profile of potential temperature and of relative humidity whose vertical variations are motivated on practical, rather than physical, grounds to be broadly representative of the range of observed soundings associated with severe weather. An ideal alternative is a model for the environmental sounding that is defined by the physics of how these environments are generated in the first place within the climate system, and whose parameters directly represent key aspects of the vertical structure of the sounding (e.g. the strength of a capping inversion). Recently, \citet[][hereafter AE17]{Agard_Emanuel_2017} developed the first theoretical model for the time-dependent one-dimensional vertical thermodynamic state associated with severe weather environments on a diurnal timescale. AE17 employs a two-layer model for the atmosphere in which the boundary layer and free troposphere may be varied independently. This state aligns with the archetypal conceptual model of the generation of high-CAPE environments east of the Rocky Mountains \citep{Carlson_Ludlam_1968,Benjamin_Carlson_1986,Benjamin_1986,Doswell_2001}. A schematic of this set-up is provided in Figure \ref{fig:conceptual}, in which warm, moist low-level air originating from the Gulf of Mexico to the south lies beneath dry well-mixed air that is advected eastward off the elevated terrain to the west. AE17 used this two-layer model framework to demonstrate analytically that peak CAPE is expected to increase with surface warming.

In principle, the AE17 theoretical model could be used to specify a steady environmental sounding for use in numerical simulation experiments. This could further allow tests of fundamental SCS sensitivities to the external physical parameters that specify the background state while holding bulk parameters (e.g. CAPE) fixed. However, this model has yet to be phrased in a way that it can directly define a sounding for use in an idealized CRM, nor has it been applied in SCS numerical simulations. Thus, there is a significant opportunity to apply this physical model for the thermodynamic environment to modern SCS numerical simulation experiments. Doing so could allow careful testing of how smaller-scale SCS morphology depends on complex variability in the vertical structure. Furthermore, given that the SCS environment represents a hydrostatic background state, this model could also be used to directly link variability in SCS soundings to the energetics of the large-scale hydrostatic atmosphere, which is the focus of modern climate physics. Such physical linkages from climate to mesoscale to storm-scale are critical for understanding both fundamental SCS environmental dependencies as well as how SCS activity may change in a future climate.

To fill this gap, this work seeks to extend AE17 to develop a novel theoretical model for a complete, steady SCS thermodynamic and kinematic sounding for use in numerical simulation experiments.
The present work focuses on how our model is constructed and provides an illustrative example of how it can be used as a theoretical foundation for both observationally motivated sensitivity testing and controlled experimentation. Thus, the specific outcomes of our simulation examples shown here are \textit{not} intended to demonstrate robust sensitivities. Moreover, the way we apply our model is by no means the only approach; it is simply a relatively straightforward one. We hope that as the model is put into use in future research it may evolve further, or perhaps it will be applied in different ways for different types of experiments. This type of comprehensive experimentation and in-depth analysis are left for future work.

Our paper is split into two parts: theory and numerical simulation. Section \ref{sec:theo} develops our theoretical sounding model and motivates the use of static energy in lieu of potential temperature as the base thermodynamic variable. An algorithm is then presented to put the model into practice, and an example comparison with the Weisman and Klemp thermodynamic model is provided to discuss similarities, differences, and benefits of our framework. Section \ref{sec:application} presents an application of how our model can be fit to a real-data sounding associated with an observed SCS event: the 3 May 1999 tornado outbreak, which was the largest outbreak in Oklahoma recorded history and produced multiple supercells and long-track tornadoes. We use this idealized sounding to demonstrate the model's potential experimental utility via illustrative sensitivity tests of variability in vertical structure at fixed CAPE and bulk shear. Finally, Section \ref{sec:conclusion} provides a summary of the model and how it may be useful for future SCS research.

\section{Theoretical model for SCS environmental sounding}\label{sec:theo}

We begin by reviewing the framework of AE17 and discuss the benefits of static energy in lieu of potential temperature for defining a hydrostatic SCS background state. Next, we develop our theoretical sounding model and an algorithm to apply it to generate an SCS sounding. Finally, we provide an example comparison with the prevailing sounding model (WK) to discuss similarities, differences, and benefits of our model framework.

\subsection{Foundation: AE17 model}

The AE17 model provides a useful foundation for generating physics-based thermodynamic environments with high CAPE amenable to SCS numerical simulation experiments. Specifically, AE17 defines the diurnal evolution of this environment with a time-dependent two-layer model for dry and moist static energies. Their idealized model begins from an initial state with constant moist static energy, where the free troposphere is dry and the boundary layer is cooler and moist; this creates convective inhibition (a capping inversion). Energy is then input into the surface at a constant rate to represent daytime solar heating, which gradually generates CAPE. AE17 used this model to test the dependence of peak CAPE on temperature on diurnal timescales.

Neglecting liquid/solid phases of water, moist static energy per unit mass $M$ is given by:
\begin{equation}\label{eq:M}
M = C_pT + L_vr + gz
\end{equation}
where $T$ is temperature, $r$ is the water vapor mixing ratio, and $z$ is geopotential height. The quantities $C_p$, $L_v$, and $g$ are the specific heat of air, the latent heat of vaporization of water, and the acceleration due to gravity, respectively, and all may be approximated as constants. Hence, moist static energy is a linear combination of temperature (sensible heat), moisture (latent heat), and altitude (potential energy). Dry static energy, $D$, is the same as $M$ but taking $r=0$, i.e.
\begin{equation}\label{eq:D}
D = C_pT + gz
\end{equation}

The AE17 model defines a thermodynamic state comprised of a free troposphere (FT) layer with constant dry static energy, $D_{FT}$, overlying a boundary layer (BL) with constant moist static energy, $M_{BL}$. The boundary layer has depth $H_{BL}$. As noted in AE17 (their Eq. 37), CAPE scales approximately with the difference between the boundary-layer moist static energy and the dry static energy, $M_{BL}-D_{FT}$, multiplied by the difference in the natural logarithm of temperatures between the level of free convection (LFC) and level of neutral buoyancy (LNB), given by
\begin{equation}
    CAPE \sim (M_{BL} - D_{FT})ln\left(\frac{T_{LFC}}{T_{LNB}}\right)
\end{equation}
Meanwhile, the difference in dry static energies between the base of the free troposphere and the boundary layer, $D_{FT}-D_{BL}$, represents a temperature jump moving upwards and hence a capping inversion. Note that this scaling neglects the effects of water vapor on buoyancy (i.e. virtual temperature effects), which will modify the true CAPE. Though not explicitly stated in AE17, the convective inhibition (CIN) follows a scaling with similar form as for CAPE, except taking the dry static energy difference across the layer bounded by the parcel level and the LFC, i.e.
\begin{equation}
    CIN \sim (D_{BL} - D_{FT})ln\left(\frac{T_{p,sfc}}{T_{LFC}}\right)
\end{equation}
where $T_{p,sfc}$ is the parcel temperature at the surface. In practice, the CIN magnitude is more strongly sensitive to neglect of moisture due to both virtual temperature effects and the effect of moisture on the height of the LCL and hence the temperature of the LFC; such errors are larger for CIN since the temperature difference across the CIN layer is relatively small compared to that across the CAPE layer.

Thus, a key benefit of the AE17 modeling framework is that CAPE and CIN may be directly modulated by varying the limited number of model physical parameters. In particular, the model explicitly incorporates an externally-defined capping inversion into the sounding. Note that a similar two-layer slab model framework is presented using potential temperature as the thermodynamic variable for understanding diurnal variability in general in \citet{Stull2012}.

\subsection{Why static energy instead of potential temperature?}

\begin{figure*}[t]
\centerline{\includegraphics[width=0.9\textwidth]{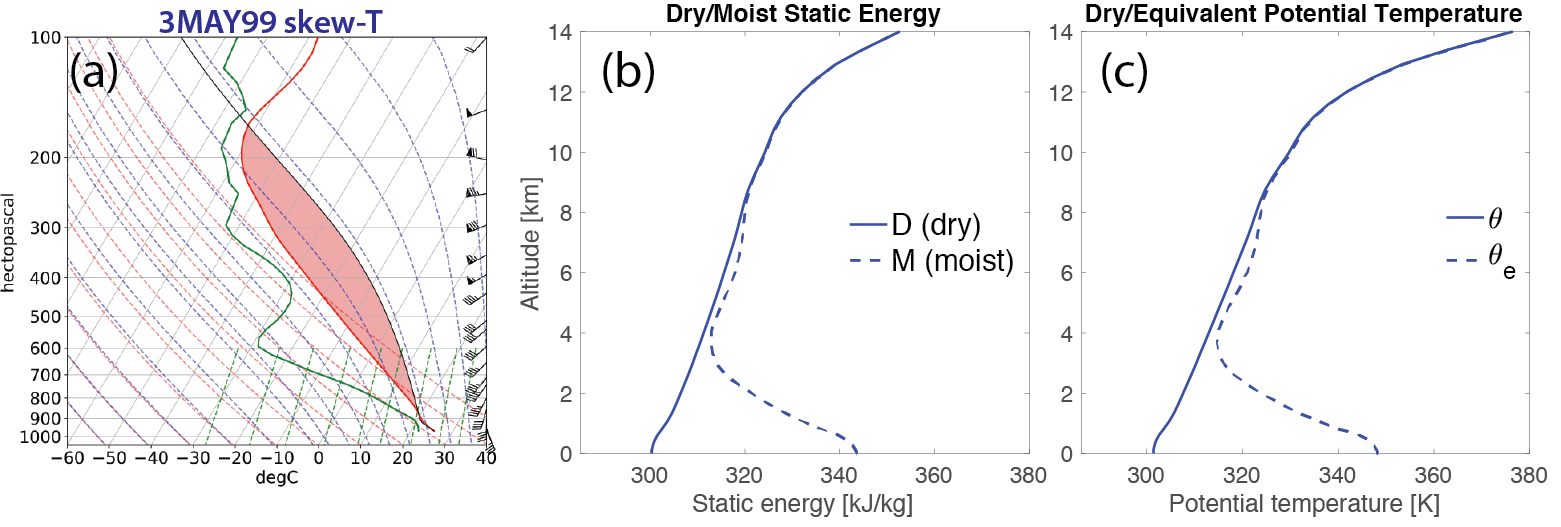}}
\vspace{-4mm}
\caption{(a) Skew-T plot for example proximity sounding from historical simulation of the 05/03/99 tornado outbreak at 2300 UTC in SW Oklahoma; (b) vertical profile of dry and moist static energies; (c) vertical profile of dry and equivalent potential temperatures.}
\label{fig:obssounding}
\vspace{-4mm}
\end{figure*}

While static energies are not commonly employed in the severe weather literature, in a hydrostatic atmosphere their vertical structures are dynamically equivalent to that of their potential temperature counterparts. For example, Figure \ref{fig:obssounding}a displays a Skew-T plot for a proximity sounding from a simulation of the 3 May 1999 tornado outbreak from \cite{DawsonII2010} (3MAY99; analyzed in detail in Section \ref{sec:application}). Figure \ref{fig:obssounding}b-c compares the vertical profiles of dry and moist static energy and dry and moist (equivalent) potential temperature for our observational case. The absolute values of these two quantities map onto one another non-linearly, but their vertical variations are very similar.

Why do potential temperature and static energy map onto one another in this way? Here we demonstrate their relationship for the dry case; the logic extends to the moist case but is significantly more complicated analytically \citep{Emanuel_2004, Bryan2008,Romps2015}. We begin from the First Law of Thermodynamics for an ideal gas, given by
\begin{equation}\label{eq:FirstLawIdeal}
    C_pdT = dq + \alpha dP
\end{equation}
where $dq$ is the external specific heating, $\alpha = \frac{1}{\rho}$ is the specific volume, and $P$ is air pressure. We then consider an adiabatic process (such as an air parcel ascending through an atmospheric column): $dq = 0$. This yields
\begin{equation}\label{eq:FirstLawAdiabatic}
    C_pdT = \alpha dP
\end{equation}

From Eq. \eqref{eq:FirstLawAdiabatic}, dry static energy requires making the assumption of hydrostatic balance,
\begin{equation}\label{eq:hydrostaticbalance}
    \alpha\frac{dP}{dz} = - g
\end{equation}
Rearranging this equation and subsituting into Eq. \eqref{eq:FirstLawAdiabatic} gives
\begin{equation}\label{eq:DALR}
    C_pdT = -gdz
\end{equation}
which can be written as the conservation equation
\begin{equation}\label{eq:dD}
    dD = 0
\end{equation}
where $D$ is the dry static energy (Eq. \eqref{eq:D})\footnote{Note that Eq. \eqref{eq:DALR} is readily rearranged to give the dry adiabatic lapse rate, which thus should formally be the ``dry adiabatic \textit{hydrostatic} lapse rate''.}. Thus, hydrostatic balance allows us to trade changes in pressure (i.e. pressure-volume work at constant pressure) with changes in altitude (i.e. potential energy).

Meanwhile, from Eq. \eqref{eq:FirstLawAdiabatic}, potential temperature requires no new assumption. Instead, we reapply the Ideal Gas Law, $P = \rho R_dT$, where $R_d$ is the specific gas constant for dry air. Rearranging this and substituting gives
\begin{equation}
    C_pd(lnT) = R_dd(lnP)
\end{equation}
which can be written as the conservation equation
\begin{equation}\label{eq:dsd}
    ds_d = 0
\end{equation}
where
\begin{equation}\label{eq:sd}
    s_d = C_plnT - R_dlnP
\end{equation}
is the dry entropy. Adding the constant $R_dlnP_0$, where $P_0$ is a reference pressure, to both sides and rearranging yields
\begin{equation}\label{eq:sdtheta}
    s_d + R_dlnP_0 = C_pln\theta
\end{equation}
where
\begin{equation}\label{eq:pottemp}
    \theta = T\left(\frac{P_0}{P}\right)^\frac{R_d}{C_p}
\end{equation}
is the dry potential temperature. We can write $\theta$ as
\begin{equation}
    \theta = e^{\frac{s_d + R_dlnP_0}{C_p}}
\end{equation}
Thus, potential temperature is an alternative, non-linear way to write entropy, in which entropy is modified by constants and then exponentiated.

How are adiabatic changes in dry entropy and dry static energy related? We start from the conservation of $s_d$ (Eq. \eqref{eq:sd}) since this requires less stringent assumptions. We use Eq. \eqref{eq:D} to write an equation for differential changes in $D$ as $C_pdT = dD - gdz$ and substitute to yield
\begin{equation}
    ds_d = \frac{1}{T}(dD - gdz) - \frac{R_d}{P}dP
\end{equation}
Reapplying hydrostatic balance, written as $- \frac{R_d}{P}dP = \frac{g}{T}dz$, gives
\begin{equation}\label{eq:entropytemp}
    ds_d = \frac{dD}{T}
\end{equation}
For hydrostatic displacements, incremental changes in entropy are simply given by incremental changes in static energy, divided by temperature. This follows from the basic thermodynamic relationship among entropy, energy, and temperature. Hence, vertical structures of entropy and static energy are qualitatively similar but differ quantitatively owing to variations in temperature with altitude. Finally, we may link $D$ to $\theta$ via Eqs. \eqref{eq:sdtheta} and \eqref{eq:entropytemp} to give
\begin{equation}\label{eq:thetasd}
    d(ln\theta) = \frac{ds_d}{C_p} = \frac{dD}{C_pT}
\end{equation}
Thus, changes in the natural logarithm of dry potential temperature are related to changes in dry static energy, normalized by the sensible heat of the parcel.

Eq. \eqref{eq:thetasd} is not very straightforward to interpret, which is the point: while potential temperature is practically useful for translating entropy to a tangible temperature-like quantity, it does so by adding non-linearity to the problem that makes it more complex analytically. Moisture further exacerbates this problem via the equivalent potential temperature ($\theta_e$), which is itself a highly non-linear combination of potential temperature and moisture. In this way, then, $\theta_e$ is remarkably useful for combining together temperature, pressure, and moisture effects into a single quantity. The downside, though, is that it makes deconstructing its components -- and the processes that control each -- much more complicated. Ultimately, while entropy is better conserved than static energy for non-hydrostatic displacements of an air parcel, such as in a thunderstorm, this more detailed accounting is not necessary for defining a hydrostatically-balanced state.

Meanwhile, static energy has practical benefits both for understanding mesoscale SCS dynamics (i.e. towards smaller scales) and for linking SCS environments to climate (i.e. towards larger scales). First, for SCS research, it is analytically simple to generate thermodynamic profiles for layers specified by dry static energy given that static energy is a linear combination of temperature, altitude, and moisture. This enables precise testing of SCS dependencies on specific aspects of the thermodynamic profile and makes it straightforward to incorporate additional modifications to the profile; an example comparison with the WK model is provided in Section \ref{sec:theo}e below. Furthermore, this framework defines the thermodynamic profile in terms of energy, which is the same physical quantity as CAPE itself; this may have useful theoretical benefits. In the end, one may readily map the model sounding back into potential temperature space as needed (e.g. in the analysis of numerical simulations) in order to work with those variables that properly account for important non-hydrostatic processes.
 
Second, for climate research, an energy-based framework offers the opportunity to directly link the hydrostatic SCS sounding to the field of climate physics, whose principal focus is the energy budget of a hydrostatic atmosphere. This budget is composed of the transfers of energy due to incoming and outgoing radiation at the top of the atmosphere, surface energy fluxes, and internal transport of energy by atmospheric and oceanic circulations \citep{Lorenz_1955,Peixoto_Oort_1992}. The partitioning of energy sources and sinks has been applied to understand variability in the global-mean climate \citep{Manabe_Strickler_1964,Meehl_1984}, horizontal variability in climate \citep{Budyko_1969,Sellers_1969,Cronin_Jansen_2016,Shaw_Barpanda_Donohoe_2018,Armour_etal_2019,Donohoe_etal_2020}, and the atmospheric response to global warming \citep{Rose_etal_2014,Roe_etal_2015,Siler_etal_2018}. Partitioning between sensible and latent heat is relevant to SCS environments given that, for example, CAPE depends on boundary layer moist static energy while CIN depends on boundary layer dry static energy as noted above. Thus, understanding how SCS activity changes with climate change requires an understanding of how the processes within the climate system alter the vertical distribution of dry and moist static energy in those hydrostatic environments that produce large values of CAPE. One great example of this is \citet{Agard_Emanuel_2017} itself, which uses an energetic framework to develop a process-level, time-dependent theory that predicts a rapid increase in peak diurnal CAPE with warming.

\subsection{Our model}

AE17 did not link their modeling framework for SCS environments to a real SCS sounding in order to be directly useful for SCS research. Our goal is to build off of the AE17 framework to develop a model for a complete, steady SCS sounding, i.e. joint thermodynamic and kinematic profiles. As described below, the model represents a transition from predominantly southerly flow advecting moist air near the surface to predominantly westerly flow advecting drier, well-mixed air aloft. In this way, the sounding is physically and intuitively consistent with the prevailing model for how severe convective storm environments are generated (Figure \ref{fig:conceptual}). A schematic of our sounding model, including both thermodynamic and kinematic profiles, is shown in Figure \ref{fig:conceptualsounding}. We explain the construction of each component next.

\subsubsection{Thermodynamic profile}

We model the thermodynamic state (Figure \ref{fig:conceptualsounding}a) beginning from the same two-layer tropospheric structure as AE17 described above: a boundary layer and a free troposphere. We then impose three additional useful modifications to put the model into practice.
\begin{enumerate}
    \item We relax the assumption of constant dry static energy in the free troposphere (i.e. dry adiabatic lapse rate, $\Gamma_d = \frac{g}{C_p}$) to allow for a constant rate of increase of dry static energy with altitude, $\beta_{FT}$. $\beta_{FT}$ sets the free-tropospheric lapse rate: from the definition of $D$, we may write $\beta_{FT} = \frac{dD_{FT}}{dz} = C_p\frac{dT_{FT}}{dz}+g$, which may be rearranged to give
    \begin{equation}
    \Gamma_{FT} = \Gamma_d - \frac{\beta_{FT}}{C_p}
    \end{equation}
    This is important given that free-tropospheric lapse rates are known to vary significantly in SCS environments \citep{Blanchard1998}. True elevated mixed layers with dry-adiabatic lapse rates are not common through the depth of the troposphere.
    \item Since AE17 does not specify a tropopause, we place a simple dry isothermal ``stratosphere'' layer with temperature $T_{tpp}$ \citep{Chavas_Emanuel_2014} at the model top whose base altitude, $H_{tpp}$, represents the tropopause altitude. $H_{tpp}$ is defined simply by the height at which the environmental temperature profile equals the tropopause temperature $T_{tpp}$. This temperature-based definition is desirable given that the tropopause temperature is expected to remain fixed locally with warming in both the tropics and midlatitudes \citep{SeeleyJeevanjeeRomps2019,HartmannLarson2002,ThompsonCeppiLi2019}.
    \item Since AE17 does not specify moisture in the free tropospheric layer, we incorporate the simplest option: constant relative humidity, $RH_{FT}(z) = RH_{FT,0}$.
\end{enumerate}

These modifications enable a more realistic representation of historical case soundings and also provide a direct means for testing variations in the thermodynamic profile at fixed CAPE. One experimental benefit of assuming constant BL dry and moist static energy is that CAPE is then insensitive to the parcel level of origin within the boundary layer. Note that the 100-mb mixed-layer CAPE (MLCAPE) is often used in forecasting because it accounts for potential boundary layer mixing by turbulence. If the assumed mixed layer extends above the top of the boundary layer, the MLCAPE may be considerably less than the surface-based CAPE (SBCAPE) depending on the magnitude of moisture near the base of the free troposphere. Because the amount of mixing depends on many details of the environment and storm evolution, we choose to focus principally on SBCAPE in this work. Additional complexities that could be added to the model, such as allowing for variations in free-tropospheric relative humidity and boundary layer moisture, are discussed below. 

\subsubsection{Kinematic profile}

\begin{figure*}[t]
\centerline{\includegraphics[width=0.9\textwidth]{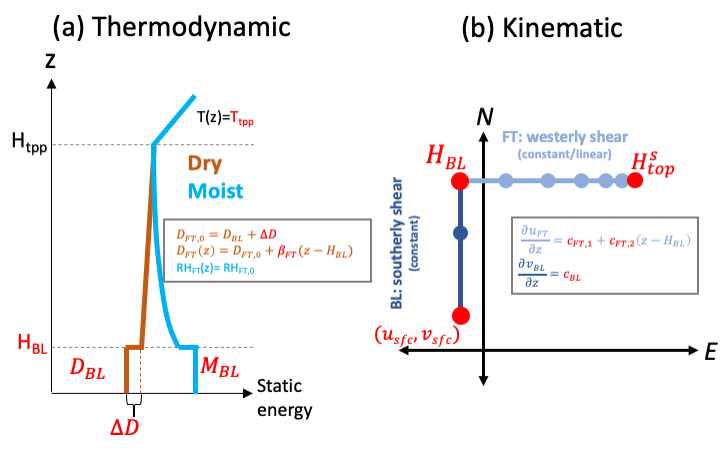}}
\caption{Schematic of model for (a) thermodynamic profile, and (b) shear profile. External parameters are colored red. The thermodynamic profile is an extension of the AE17 model. The kinematic model assumes constant southerly shear in the boundary-layer, constant or linearly-decreasing westerly shear in the free troposphere, and sets the boundary layer height equal to its value in the thermodynamic profile.}
\label{fig:conceptualsounding}
\end{figure*}

A schematic of the model kinematic profile is shown in Figure \ref{fig:conceptualsounding}b. We propose a similar two-layer model for representing the kinematic structure of the sounding that is physically consistent with the thermodynamic model. Our model is similar to recent work idealizing the kinematic profile using L-shaped hodographs that are often seen in tornadic supercell environments \citep{Esterheld2008,Beck2013,Sherburn2015,Guarriello2018,Petersetal2020}. The model is comprised of a lower free-tropospheric layer superimposed over a boundary layer with the same depth as the thermodynamic model ($H_{BL}$). Each layer is assumed to have unidirectional shear, with the boundary layer defined relative to a specified surface wind, $(u_{sfc},v_{sfc})$.

The boundary shear layer is specified with constant southerly shear, i.e.
\begin{align}
    \frac{\partial u_{BL}}{\partial z} &= 0 \\
    \frac{\partial v_{BL}}{\partial z} &= c_{BL}
\end{align}
$c_{BL}$ represents the constant meridional shear in the boundary layer. The bulk vector shear across the boundary layer is thus:
\begin{equation}
    \Delta V_{BL} = \int_{0}^{H_{BL}} \frac{\partial v_{BL}}{\partial z} dz = c_{BL}H_{BL}
\end{equation}

The upper shear layer extends from the base of the free troposphere up to a fixed altitude, $H^s_{top}$. The layer is specified with westerly shear. We allow this shear to be constant or linearly-decreasing with height, as shear is often concentrated at lower levels in convective storm environments, particularly those associated with tornadic supercells \citep[e.g., ][]{Esterheld2008,Coffer2015,Thompson2003,Coffer2019}, i.e.
\begin{align}
    \frac{\partial u_{FT}}{\partial z}(z) &= c_{FT,1} + c_{FT,2}(z-H_{BL}) \label{eq:shear_FT_u} \\
    \frac{\partial v_{FT}}{\partial z} &= 0
\end{align}
where $c_{FT,1}$ represents the zonal shear at the base of the upper shear layer and $c_{FT,2}$ represents the rate of change of zonal shear with height. Eq. \eqref{eq:shear_FT_u} represents the transition from a zonal shear magnitude of $c_{FT,1}$ at the layer base ($z=H_{BL}$) to $c_{FT,1}+c_{FT,2}(H^s_{top}-H_{BL})$ at the layer top ($z=H^s_{top}$). The bulk vector shear across the upper shear layer is thus:
\begin{equation}
    \Delta V_{FT} = \int_{H_{BL}}^{H^s_{top}} \frac{\partial u_{FT}}{\partial z} dz = c_{FT,1}(H^s_{top} - H_{BL}) + \frac{1}{2}c_{FT,2}(H^s_{top} - H_{BL})^2
\end{equation}
Thus for a fixed value of bulk layer shear, a range of combinations of $(c_{FT,1},c_{FT,2})$ are possible. There are two simple limit cases to consider for the upper shear layer:
\begin{enumerate}
    \item Constant shear: $c_{FT,2}=0$; the zonal shear $\frac{\partial u_{FT}}{\partial z}(z) = c_{FT,1}$ is constant throughout the layer
    \item Shear decreasing linearly to zero at the layer top: $c_{FT,2}=-c_{FT,1}/(H^s_{top}-H_{BL})$; the zonal shear (Eq. \eqref{eq:shear_FT_u}) reduces to $\frac{\partial u_{FT}}{\partial z}(z) = c_{FT,1}\left(1 - \frac{z-H_{BL}}{H^s_{top}-H_{BL}}\right)$ .
\end{enumerate}

For the remainder of the shear profile ($z>H^s_{top}$), we impose zero shear (i.e. constant wind vector).

\subsection{Practical implementation of model}

Our objective is to use the model sounding in numerical simulations. We define our model moving upwards from the surface, similar to how a sounding is obtained by an ascending radiosonde.

\subsubsection{Thermodynamic profile}

The most straightforward implementation of the thermodynamic model is as follows:
\begin{enumerate}
    \item Calculate surface dry and moist static energy, $D_{sfc}$ (Eq. \eqref{eq:D}) and $M_{sfc}$ (Eq. \eqref{eq:M}), from input surface pressure, temperature, and relative humidity $(P_{sfc},T_{sfc},RH_{sfc})$.
    \item Calculate temperature in the boundary layer ($z \le H_{BL}$) assuming constant dry static energy, $D_{BL}=D_{sfc}$: $T_{BL}(z) = \frac{1}{C_p}(D_{BL}-gz)$.
    \item Calculate mixing ratio in the boundary layer ($z \le H_{BL}$) assuming constant moist static energy, $M_{BL}=M_{sfc}$. This translates simply to holding mixing ratio constant: $r_{BL}(z) = r_{sfc}$ (well-mixed). Mixing ratios are capped such that relative humidity does not exceed 99\% (note: this is performed in the final step and thus reduces $M_{BL}$ at those levels).
    \item Calculate dry static energy at the base of the free troposphere, defined as the first level above $H_{BL}$: $D_{FT,0} = D_{BL}(H_{BL}) + \Delta D$
    \item Calculate dry static energy in the free troposphere ($z>H_{BL}$): $D_{FT}(z) = D_{FT,0} + \beta_{FT} (z-H_{BL})$. This quantity defines the free-tropospheric lapse rate, $\Gamma_{FT} = \Gamma_d - \frac{\beta_{FT}}{C_p}$.
    \item Calculate temperature in the free troposphere from $D_{FT}(z)$: $T_{FT}(z) = \frac{1}{C_p}(D_{FT}(z)-gz)$.
    \item Integrate hydrostatic balance (Eq. \eqref{eq:hydrostaticbalance}) upwards from the surface pressure $P_{sfc}$ to calculate the hydrostatic pressure at all altitudes and the mixing ratio in the free troposphere\footnote{Technically these two integrations should be repeated until they converge to account jointly for the hydrostatic pressure of the free tropospheric moisture overhead and the pressure dependence of the mixing ratio, though the errors are generally very small for Earth-like temperatures.}. Mixing ratio is calculated using: $r = \epsilon\frac{(RH)e^*}{P-(RH)e^*}$, where $e^*$ is the saturation vapor pressure and $RH_{FT}(z) = RH_{FT,0}$.
    \item Impose a dry isothermal ``stratosphere" (i.e. statically-stable) at the model top. This is done by setting the temperature to $T_{tpp}$ and the mixing ratio to zero at all altitudes where the predicted free-tropospheric temperature from the previous step is less than the tropopause temperature, $T<T_{tpp}$.
\end{enumerate}

This algorithm specifies the thermodynamic model from the following eight external parameters: $P_{sfc}$, $T_{sfc}$, $RH_{sfc}$, $H_{BL}$, $\Delta D$, $\beta_{FT}$, $RH_{FT,0}$, and $T_{tpp}$.

\subsubsection{Kinematic profile}

The shear profile may be similarly defined moving upwards from the surface:
\begin{enumerate}
    \item Define the input surface wind vector, $(u_{sfc},v_{sfc})$.
    \item Calculate the boundary shear layer flow velocities ($z \le H_{BL}$): $u_{BL}(z) = u_{sfc}$, $v_{BL}(z) = v_{sfc}  + c_{BL}z$.
    \item Calculate the upper shear layer flow velocities ($H_{BL} < z \le H^s_{top}$): $u_{FT}(z) = u_{BL}(H_{BL}) + c_{FT,1}(z-H_{BL}) + \frac{1}{2}c_{FT,2}(z-H_{BL})^2$, $v_{FT}(z) = v_{BL}(H_{BL})$.
    \item Set flow velocities constant for $z>H^s_{top}$: $u(z)=u_{FT}(H^s_{top})$, $v(z)=v_{FT}(H^s_{top})$
\end{enumerate}

This algorithm specifies the kinematic model from the following six external parameters: $u_{sfc}$, $v_{sfc}$, $c_{BL}$, $c_{FT,1}$, $c_{FT,2}$, and $H^s_{top}$. $H_{BL}$ is defined in the thermodynamic model.

\subsubsection{Summary and additional potential modifications}

The above is a minimal theoretical model that contains the necessary ingredients for a viable environmental SCS sounding -- i.e. one with significant CAPE and vertical wind shear, and relatively low CIN. We emphasize here that this does \textit{not} guarantee that a given sounding specified by this model will produce any specific SCS outcome, such as a long-lived supercell. In this way, then, the base model provides a natural starting point for testing how changes to the sounding affect SCS outcomes, as demonstrated in Section \ref{sec:application}.

We have incorporated a few additional types of complexity to better capture real-world soundings. Without question, there are numerous additional degrees of complexity that could be readily added to the model to test their significance. We highlight a few possible options here:
\begin{itemize}
\item Relaxing the constant moist static energy constraint in the boundary layer to allow for representation of moisture entrainment or detrainment from the free troposphere. \citet{Schultz2012} found that significant tornadoes from discrete supercells were more likely when the boundary layer was capped and the $\theta_e$ (and hence moist static energy) was constant or \textit{increased} with height. Note that this will introduce new variation in CAPE and CIN calculated for parcels from different levels (or vertically-averaged) within the BL.
\item A water vapor or relative humidity lapse rate at the base of the free troposphere, to allow for a more gradual moisture transition across the capping inversion. In our model, this transition is sharp.
\item Multiple free-tropospheric layers. For example, here we have allowed free tropospheric moisture to vary independently of temperature (dry static energy), which is not characteristic of a true EML. A real EML would also have constant mixing ratio, since the EML was once a well-mixed boundary layer itself. Such a layer could be applied as an intermediate layer in the lower free-troposphere.
\item Height dependence of shear in the boundary layer. Recent studies have found evidence that strong shear in the lowest few hundred m AGL is more closely related to significant tornado occurrence in supercell storms than the 0-1 km layer more commonly utilized in operational contexts \citep{Markowski2003,Esterheld2008,Coffer2019}.
\end{itemize}

\subsection{Comparison with Weisman and Klemp sounding}

The WK sounding is characterized by simple, smoothly-varying analytic expressions for potential temperature and relative humidity as a function of altitude. Potential temperature increases from a specified surface value to a specified tropopause value and then increases exponentially above the tropopause, according to
\begin{equation}\label{eq:theta_WK}
\theta(z) =
    \begin{cases}
      \theta_{sfc}+(\theta_{tpp}-\theta_{sfc})\left(\frac{z}{z_{tpp}}\right)^\frac{5}{4}, & \text{if}\ z \le z_{tpp} \\
      \theta_{tpp} exp\left[\frac{g(z-z_{tpp})}{C_pT_{tpp}}\right], & \text{if}\ z > z_{tpp}
    \end{cases}
\end{equation}
The latter equation yields an isothermal layer above the tropopause (shown analytically in Appendix A; cf. WK Figure 1), which is identical to our model. Note that the tropopause is overspecified in this formulation -- its height, temperature, and potential temperature are all input parameters. As a result, changing the value of $T_{tpp}$ alone does not actually alter the tropopause in the same manner in the profile itself; one must first solve for one parameter from the solution below the tropopause before specifying the solution above the tropopause.

Relative humidity decreases moving upwards according to a similar dependence on altitude
\begin{equation}
RH(z) = 1-\frac{3}{4}\left(\frac{z}{z_{tpp}}\right)^\frac{5}{4}
\end{equation}
and is set constant at 0.25 above the tropopause. Finally, a boundary layer that is well-mixed in moisture is created by imposing an upper-bound on the water vapor mixing ratio, $r_{v,sfc}$, which reduces the RH at all levels where the initial $r_v$ value exceeds $r_{v,sfc}$. The boundary layer is not well-mixed in potential temperature.

\begin{figure*}[t]
\centerline{\includegraphics[width=0.9\textwidth]{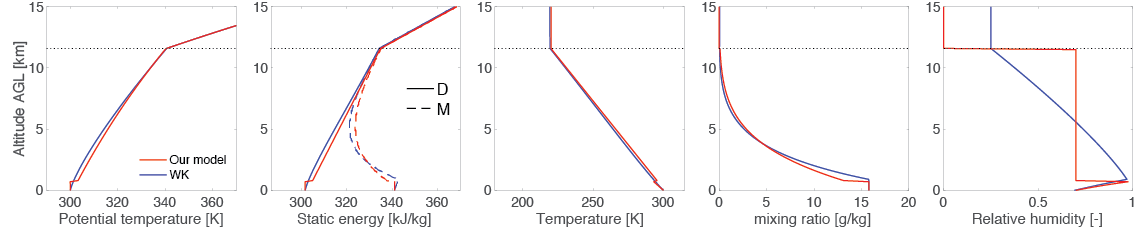}}
\caption{Example of our model thermodynamic state (red) and comparison with the WK model (blue). WK input parameters are defined directly from our model sounding.}
\label{fig:WKcompare}
\end{figure*}

Figure \ref{fig:WKcompare} displays an example of our model thermodynamic profile with comparison against WK. The input parameters for WK are defined directly from our model sounding. This comparison allows us to highlight similarities and differences in model construction. The parameters for our model are: $P_{sfc} = 1000 \: hPa$, $T_{sfc} = 300 \: K$, $RH_{sfc} = 0.7$, $H_{BL} = 700 \: m$, $\Delta D = 3000 \: J/kg$, $T_{tpp} = 220 \: K$, $\Gamma_{FT} = 7.0 \: K/km$, and $RH_{FT,0} = 0.7$. The resulting parameters for WK are: $\theta_{sfc} = 300 \: K$, $\theta_{tpp} = 340.6 \: K$, $z_{tpp} = 11.6 \: km$, and $r_{v,sfc} = 15.8 \: g/kg$; $T_{tpp}$ and $P_{sfc}$ are the same as above.

The thermal profiles are overall quite similar. Note that the $\frac{5}{4}$ exponent used in WK for the increase in potential temperature with height yields a free-tropospheric lapse rate that is relatively close to constant; our model imposes this structure by definition. The principal difference is the existence of an explicit, sharp capping inversion in our model, which is a result of the two-layer tropospheric framework. Such a sharp inversion is not straightforward to produce in WK owing to its simpler construction \citep{Naylor_Askelson_Gilmore_2012}.

The relative humidity profiles are nearly identical within their respective boundary layers. Note that the boundary layer depth in our model may be varied independently of $r_{v,sfc}$, whereas in WK the two are intrinsically linked. In the free troposphere, WK again imposes a $\frac{5}{4}$ exponent, which yields an RH profile that decreases quasi-linearly with altitude. In contrast, our model simply assumes constant RH, though a linear decrease with altitude could readily be added. Neither choice is ``correct'' nor more physical than the other. Arguably the most logical structure based on observations is a "C-shaped" profile, as relative humidity is generally high at in the boundary layer and near the tropopause with a local minimum in the middle free troposphere \citep{Gettelmanetal2006,Romps2014}. Ultimately, though, free tropospheric RH is poorly constrained for SCS research given that CAPE for a boundary layer parcel is relatively insensitive to free tropospheric moisture. Hence it is left constant in our model for simplicity, which may serve as a baseline for comparison with more complex vertical structures.

Overall, our model offers useful physical insight into the vertical structure of the WK model, whose parametric formulation was motivated by a practical need to represent real-world soundings. Our model more explicitly represents key aspects of this vertical structure:
\begin{itemize}
\item A distinct boundary layer and free troposphere whose properties (temperature and moisture) can be varied independently;
\item A capping inversion (as represented by the dry static energy jump between the two tropospheric layers);
\item A well-mixed boundary layer;
\item Direct specification of the free tropospheric lapse rate, in lieu of the arbitrary $\frac{5}{4}$ power law increase in $\theta$ with height
\end{itemize}

We note that there may be experimental applications for which the WK model is equally viable for defining a sounding or set of soundings. At a minimum, our model can provide clearer physical motivation for the structure of any idealized sounding. Our model can otherwise offer more precise control over the structure of the sounding (both thermodynamic and kinematic) and its relationship to key quantities, such as CAPE and CIN, via its physical parameters. Finally, the use of static energy is consistent with CAPE as an energy quantity as well as with the large-scale energetics of a hydrostatic atmosphere as noted earlier.






\section{Application to historical case: the 3 May 1999 tornado outbreak}\label{sec:application}

We next provide a demonstration of how our model may be used to idealize an SCS environmental sounding associated with a real historical event. We then demonstrate the experimental utility of the model via illustrative sensitivity tests of variability in vertical structure in SCS numerical simulations.

\subsection{Numerical simulation description}

Experiments are performed using the CM1 numerical model \citep{Bryan_Fritsch_2002} version 19. CM1 is a fully compressible nonhydrostatic computational model designed for idealized simulations of mesoscale and smaller atmospheric phenomena. CM1 has been employed to gain fundamental insight into a wide range of mesoscale phenomena in both the mid-latitudes and tropics, including severe convective storms and tornadoes \citep{Bryan2006, James2010, Naylor2012, Orf2017, Dahl2012, Dahl2014,Naylor2014,Parker2014,Dahl2015,Markowski2016,Peters2016,PetersHannahMorrison2019}, supercells \citep{James2010,Coffer2015,Davenport2015,Nowotarski2016} and convective squall lines \citep{Bryan2006}; tropical cyclones \citep{Bryan_Rotunno_2009,Chavas_Emanuel_2014,Davis_2015,Navarro_Hakim_2016,Naylor_Schecter_2014,Bu_Fovell_Corbosiero_2014,Peng_Rotunno_Bryan_2018}; and the scaling of vertical velocity, precipitation extremes, and CAPE with climate in radiative-convective equilibrium \citep{Singh_OGorman_2013,Singh_OGorman_2014,Singh_OGorman_2015}.

CM1 is particularly well-suited for this work for a number of reasons, including 1) it has demonstrated flexibility across a range of scales and scientific questions; 2) its excellent mass and momentum conservation properties; and 3) inclusion of various thermodynamic terms often neglected in other numerical models (such as the heat capacity of hydrometeors), which may be important on convection-resolving scales. Moreover, CM1 uses a height-based vertical coordinate, which fits naturally with our static energy-based theoretical sounding framework.


\subsection{Experiments}

The setup of the simulation domain, grid parameters, and physical parameterizations closely follows that of \cite{DawsonII2019}, though we neglect the Coriolis force and use free-slip lower boundary conditions only. Each of our simulation experiments is performed on a 200 km $\times$ 200 km $\times$ 20 km domain with a horizontal grid spacing of 250 m in an inner 100 $\times$ 100 km$^2$ region and gradually stretched to 1 km at the lateral boundaries. The lateral boundary conditions are open radiative, while the top and bottom boundaries are impermeable and free-slip. A Rayleigh damping layer is located above 15 km with an inverse e-folding time of 1/300 s$^{-1}$. The vertical grid has 50 levels stretched from 20 m at the surface to $\sim$800 m at the domain top (20 km). The domain translates with a constant [u, v] = [7.28, 8.78] m s$^{-1}$ to keep the simulated storm near the center of the domain. Deep convection is initiated using the \citet{Naylor2012} updraft nudging technique applied to an ellipsoidal region with maximum $w=$10 m s$^{-1}$ and radii 10 km $\times$ 10 km $\times$ 1.5 km and centered at [x, y, z] = [100, 100, 1.5] km over the first 900 s of numerical model integration. The NSSL triple-moment microphysics scheme \citep{Mansell2010, DawsonII2014} and a 1.5-order prognostic TKE turbulence closure scheme \citep{Deardorff1980} is used. Finally, as is common in idealized CRM simulations of deep convection, no radiation or surface physics are included. All simulations are run for 4 h.
We perform simulation experiments using four soundings described in Table 1 to define the horizontally homogeneous initial environment.

We first perform a simulation with our example historical event sounding (3MAY99) from \cite{DawsonII2010} shown in Figure \ref{fig:obssounding}. We then perform simulations with our model fit to 3MAY99 (THEO; fitting described in Section \ref{sec:application}\ref{subsec:soundingfit}). Finally, we perform two experiments to illustrate distinct uses of the model: 1) experiment MODHIST, which uses a semi-theoretical sounding that tests the inclusion of specific details of the real sounding into the theoretical model (here: low-level moisture); and 2) experiment MODTHEO, which uses a fully-theoretical sounding that tests direct modifications of theoretical model parameters (here: free-tropospheric relative humidity).

\begin{table}[t!]
\centering
\resizebox{0.5\textwidth}{!}{%
\begin{tabular}{|p{1.0in}|p{2.6in}|}
\hline
    \underline{Name} & \underline{Details} \\
    \hline
    3MAY99 (Historical) & Proximity sounding from a simulation of the 3 May 1999 tornado outbreak, from \cite{DawsonII2010} \\
    \hline
    THEO & Pure theoretical model fit to 3MAY99 \\
    \hline
    MODHIST & THEO with the low-level moisture set equal to values from 3MAY99 historical event sounding ($z \le 0.84 \: km$)  \\
    \hline
    MODTHEO &  THEO with enhanced constant free tropospheric relative humidity (70\%) \\
    \hline
\end{tabular}%
}
\caption{Soundings for our experiments.}
\label{tab:1}
\end{table}

\subsection{Fitting the model sounding}\label{subsec:soundingfit}

We fit our model thermodynamic profile to the historical event sounding as follows:
\begin{itemize}
    \item Set $P_{sfc}$, $T_{sfc}$, $RH_{sfc}$ equal to the observed values.
    \item Set $H_{BL}$ equal to the level of maximum $RH$.
    \item Set $\Delta D$ equal to the difference between the mean dry static energy in $z \in (H_{BL}, \; 3H_{BL}]$ and the mean dry static energy in $z \le H_{BL}$. This captures the enhanced dry static energy at the base of the free troposphere.
    \item Set $T_{tpp}$ equal to the coldest temperature in the sounding, whose altitude defines $H_{tpp}$.
    \item Set $\Gamma_{FT}$ equal to the mean lapse rate in the layer $z \in [H_{BL}, \; H_{BL} + 0.75(H_{tpp}-H_{BL})]$. This average avoids the top of the troposphere where lapse rates necessarily become more stable as they approach the tropopause.
    \item Set $RH_{FT,0}$ equal to the free-tropospheric column saturation fraction, $\frac{W}{W^*}$. The quantity
    $W = \int_{z_{bot}}^{z_{top}}\rho q_v dz' = \frac{1}{g}\int_{P_{bot}}^{P_{top}}q_v dP'$ is the water vapor path and $W^*$ is its saturation value \citep{BrethertonPetersBack2004,Camargo_etal_2014,RaymondSessionsFuchs2007}. Each term is calculated within the free-tropospheric layer $z\in(H_{BL},H_{tpp})$. Saturation fraction is effectively identical to a mass-weighted relative humidity (see Appendix B), and hence is also commonly called column relative humidity. This approach yields a sounding with nearly the same water vapor path as exists in the real sounding and thus avoids the addition of significant artificial sources or sinks of latent heat into the column.
\end{itemize}
This algorithm will yield values of SBCAPE similar to the historical event sounding.

\begin{figure*}[t]
\centerline{\includegraphics[width=0.9\textwidth]{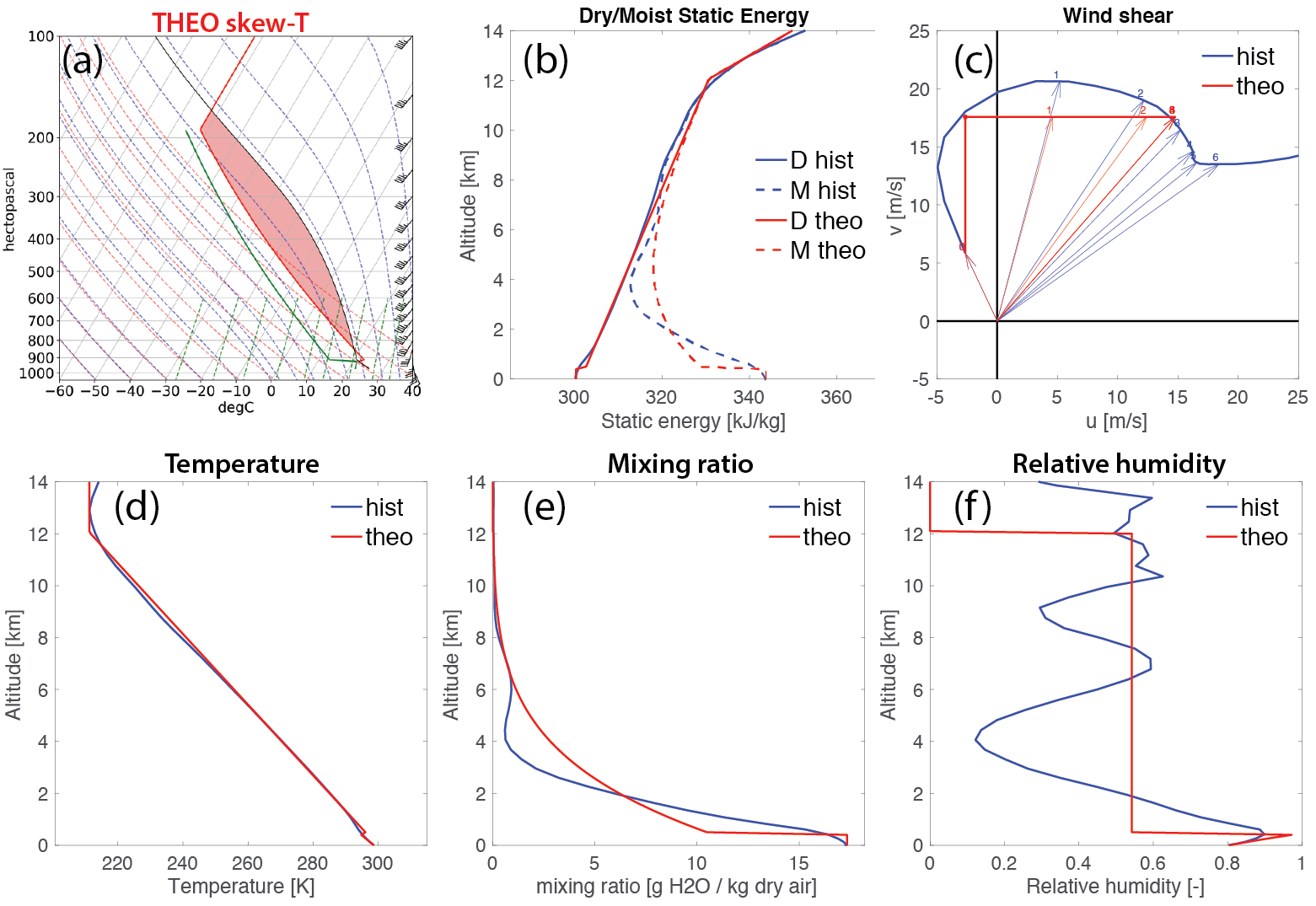}}
\vspace{-4mm}
\caption{Comparison of soundings for 3MAY99 historical event (blue) and THEO (red). (a) THEO Skew-T; (b) dry and moist static energies; (c) wind shear; (d) temperature; (e) mixing ratio; (f) relative humidity. Both soundings have similar surface-based CAPE (3MAY99: $4711 \; J \; kg^{-1}$; THEO: $4490 \; J \; kg^{-1}$) and identical 0-3 km bulk shear ($21 \: m \: s^{-1}$).}
\label{fig:obsfitssounding}
\vspace{-4mm}
\end{figure*}

We fit our model kinematic profile to the sounding as follows:
\begin{itemize}
    \item Set $(u_{sfc},v_{sfc})$ equal to the observed values.
    \item Set $c_{BL} = \left|\overline{\frac{\partial \textbf{V}_{BL}}{\partial z}}\right|$ equal to the average vector shear magnitude in $z<H_{BL}$. This matches the bulk total shear magnitude between the surface and $H_{BL}$ and distributes this shear purely in the southerly direction.
    \item Set $H^s_{top}$ to 3 km. This focuses on the low-level shear; in the 3MAY99 sounding, most of the shear is confined to below 3 km (Figure \ref{fig:obsfitssounding}c).
    \item Set $c_{FT,1} = 2\left|\overline{\frac{\partial \textbf{V}_{FT}}{\partial z}}\right|$, where $\left|\overline{\frac{\partial \textbf{V}_{FT}}{\partial z}}\right|$ equals the average vector shear magnitude in $H_{BL}<z<H^s_{top}$, and set $c_{FT,2}=-c_{FT,1}/(H^s_{top}-H_{BL})$. This combination matches the bulk total shear magnitude between $H_{BL}$ and $H^s_{top}$ and distributes this shear purely in the westerly direction, with a magnitude that decreases linearly to zero at $z=H^s_{top}$ (as noted above).
\end{itemize}
This algorithm also matches the total bulk shear in the sounding across \textit{both} layers ($z \le H_s^{top}$). One potential additional kinematic constraint would be to fit the shear profile to the storm-relative helicity. However, this requires precise knowledge of the storm-motion vector, which is a complex function of both the wind profile and internal storm processes such as cold pool propagation \citep[e.g., ][]{Bunkers2018}. Nonetheless, we think this could be a valuable addition that we leave for future work.

Our approach is certainly not the only way to fit the model parameters. For example, while we fit the 0-3 km shear in this study, we note that the 0-6 km is the standard shear layer for SCS forecasting \citep{Doswell_2001}. However, the model can be fit to any shear layer depending on the experimental purpose.


Figure \ref{fig:obsfitssounding} displays the theoretical sounding (red, THEO) fit to our example historical event sounding (blue, 3MAY99) following the algorithm described above. For the THEO thermodynamic profile (Figure \ref{fig:obsfitssounding}b), the boundary layer dry and moist static energies equal their respective 3MAY99 near-surface values. In the free troposphere, the dry static energy jump is $\Delta D = 2095 \ J/kg$; the boundary layer height is $H_{BL} = 0.42 \; km$; the relative humidity is $RH_{FT,0} = 0.54$; the free-tropospheric lapse rate is $\Gamma_{FT} = 7.34 \; K/km$; and the tropopause temperature is $T_{tpp} = 211.25 \; K$. For the THEO kinematic profile (Figure \ref{fig:obsfitssounding}c), the surface flow vector equals the 3MAY99 value of $(u_{sfc},v_{sfc}) = (-2.64, 5.83) \; m \: s^{-1}$. The shear profile constants are $c_{BL} = 0.0293 \; s^{-1}$, $c_{FT,1} = 0.0139 \; s^{-1}$, and $c_{FT,2} = -5.367*10^{-6} \; m^{-1}s^{-1}$. Both soundings have similar surface-based CAPE: $4711 \; J \; kg^{-1}$ for 3MAY99 and $4490 \; J \; kg^{-1}$ for THEO. Both soundings have identical 0-3 km bulk shear of $21 \; m \: s^{-1}$.

To compare the profiles in terms of standard meteorological variables, Figure \ref{fig:obsfitssounding}d-f compare temperature, mixing ratio, and relative humidity between THEO and 3MAY99. The temperature structure is remarkably similar at all levels except near the top of the boundary layer where THEO has a sharper capping inversion, indicating that in this case the use of a single free-tropospheric lapse rate is quite reasonable. Meanwhile, clearly there are significant vertical variations in free-tropospheric moisture in 3MAY99 that are not represented in our simple model, including greater moisture in the lower free-troposphere and less moisture in the middle free-troposphere. The role of these detailed variations could be tested in future experiments.

\subsection{SCS simulation experiments}\label{subsec:simulation}

\begin{figure*}[t]
\vspace{-2mm}
\centerline{\includegraphics[width=\textwidth]{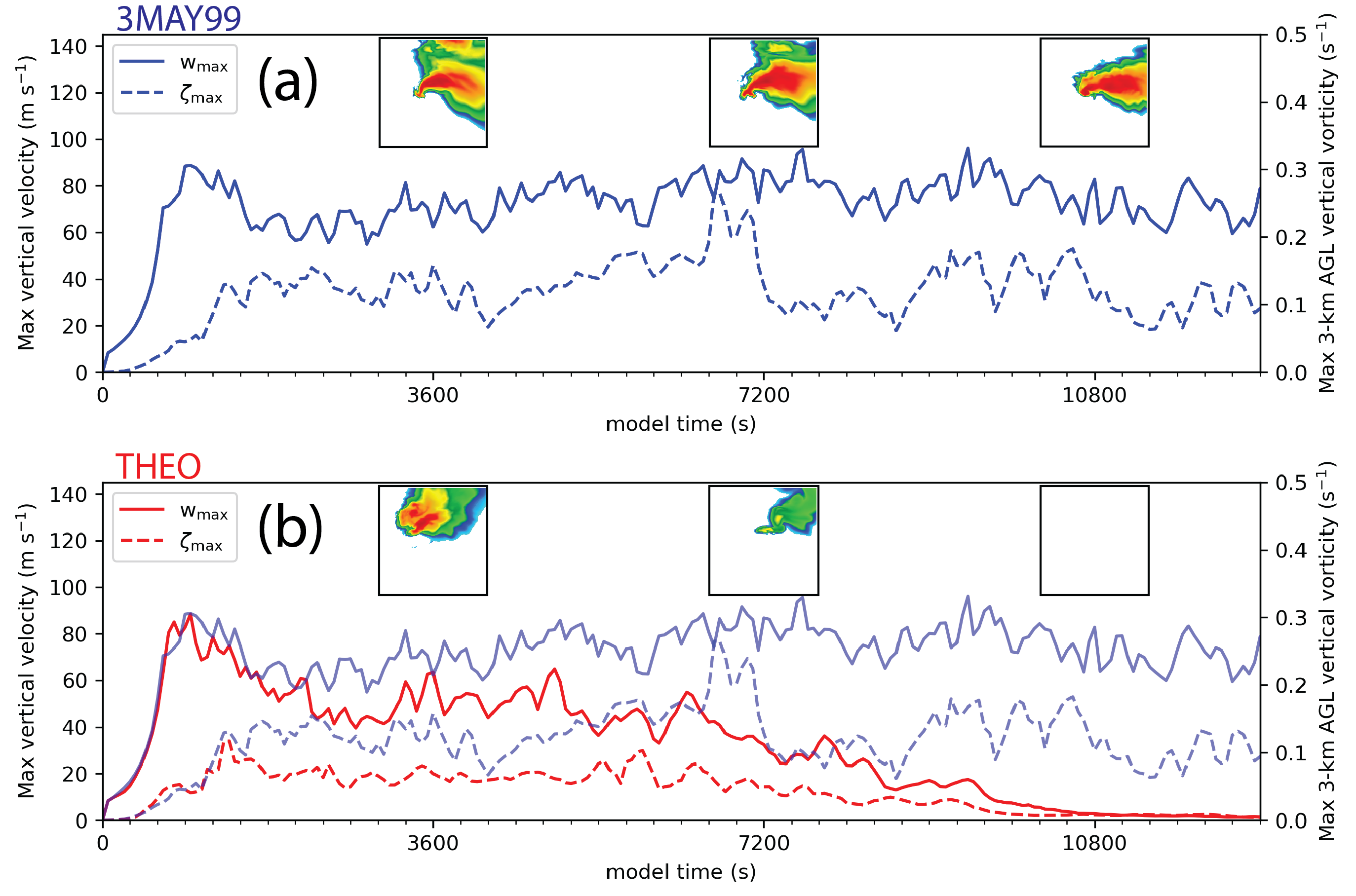}}
\vspace{-4mm}
\caption{Simulated supercell evolution associated with (a) historical-event sounding (3MAY99) shown in Figure \ref{fig:obssounding} from the 3 May 1999 tornado outbreak; (b) theoretical sounding (THEO; red) fit to the 3MAY99 sounding, with 3MAY99 result (blue) repeated for comparison.   Timeseries show peak vertical velocity (solid) and 3km AGL vertical vorticity (dash) at 3 km AGL, with snapshots of reflectivity (dBz, boxes).}
\label{fig:obstheosimulation}
\vspace{-4mm}
\end{figure*}

We first perform a numerical simulation experiment using the 3MAY99 sounding. Figure \ref{fig:obstheosimulation}a displays time series of domain maximum vertical velocity and maximum vertical vorticity at 3km AGL and snapshots of surface simulated radar reflectivity at 1, 2, and 3 h into the simulation. Our 3MAY99 simulation successfully produces a long-lived supercell. Next, we perform a simulation experiment using the theoretical sounding (THEO) and compare against 3MAY99, as shown in Figure \ref{fig:obstheosimulation}b. While 3MAY99 produces a long-lived supercell, THEO yields a short-lived convective cell that quickly dissipates after 1 hour despite having environments with similar SBCAPE and 0-3 km bulk shear.

\begin{figure*}[t]
\centerline{\includegraphics[width=0.9\textwidth]{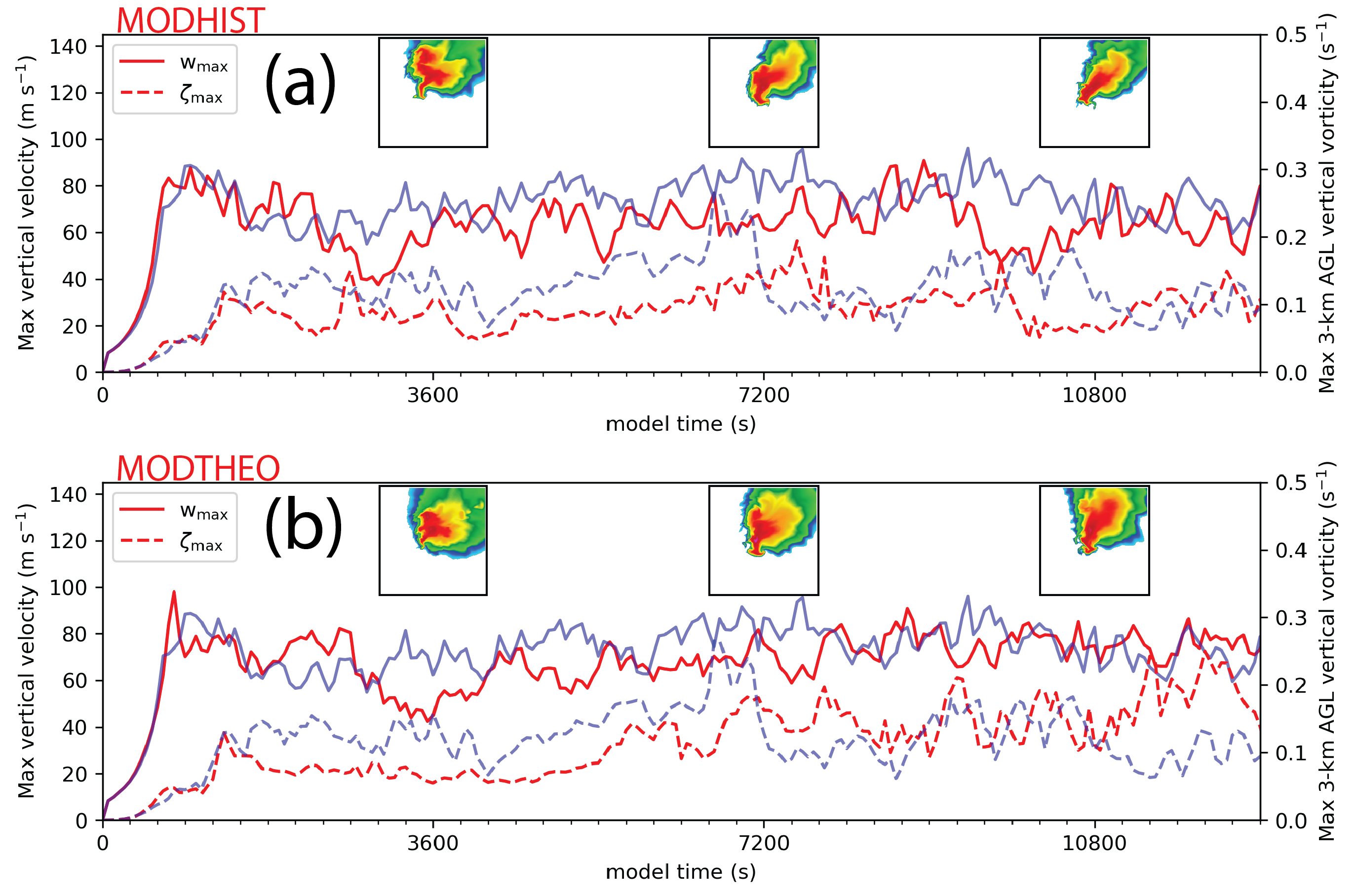}}
\vspace{-4mm}
\caption{Simulated supercell evolution for two experiments modifying THEO. (a) MODHIST (red), whose sounding is identical to that of THEO except with $r_v$ forced to match 3MAY99 in the lowest 0.84 km. (b) MODTHEO (red), whose sounding is identical to that of THEO except with $RH_{FT,0}$ enhanced to 70\%. 3MAY99 evolution also shown (blue). In both experiments, a long-lived supercell emerges as was found with 3MAY99. Plot aesthetics as in Figure \ref{fig:obstheosimulation}b.}
\label{fig:modexperiments}
\end{figure*}

Finally, we perform two demonstration experiments in which we modify our THEO sounding to illustrate the experimental utility of our model. The sounding used in the first experiment (MODHIST) is "semi-theoretical" in that it is identical to THEO but in which $r_v$ is forced to match 3MAY99 in the lowest 0.84 km (i.e. $2H_{BL}$). Thus it demonstrates how a specific feature of a real-data sounding may be incorporated into the model to test its importance. The result is shown in Figure \ref{fig:modexperiments}a. The experiment with this slight modification now produces a long-lived supercell. The next experiment (MODTHEO) is ``fully-theoretical'' and demonstrates how physical parameters in the theoretical model can be directly varied to test their importance. Experiment MODTHEO is identical to THEO but with the free-tropospheric relative humidity, $RH_{FT,0}$, enhanced to 70\%. The result is shown in Figure \ref{fig:modexperiments}b. This experiment also produces a long-lived supercell, similar to both 3MAY99 and MODHIST.

Overall, the results across our experiments suggest a substantial sensitivity of convective evolution to the vertical structure of moisture in both the BL and FT. They are consistent with many past studies that highlight the important role of variability in the vertical thermodynamic structure in governing storm dynamics \citep[e.g., ][]{McCaul_Weisman_2001,McCaul2002,McCaul2005,Cohen2007,Kirkpatrick2009,James2010,DawsonII2012,Guarriello2018,Brown2019}. For example, we note that moisture at the base of the free troposphere varies across our experiments and hence yields different values of MLCAPE and MLCIN. Owing to both the shallower moisture and the sharper cap in THEO, the MLCAPE and CIN for THEO is 2447 J kg$^{-1}$ and -47 J kg$^{-1}$, respectively, as compared with 3852 J kg$^{-1}$ and -6 J kg$^{-1}$ for 3MAY99. For MODHIST, by replacing the THEO low-level moisture profile with that in 3MAY99, the MLCAPE increases to 3552 J kg$^{-1}$ and the MLCIN decreases to -15 J kg$^{-1}$, while the SBCAPE and SBCIN are very similar to 3MAY99. For MODTHEO, despite having the same BL structure as THEO, the increase in free-tropospheric moisture causes the MLCAPE to increase to 3370 J kg$^{-1}$ and the MLCIN to decrease to -19 J kg$^{-1}$. Thus, one hypothesis for the failure to produce a long-lived supercell in THEO is that at least some updraft source parcels for the simulated storm are coming from above the boundary layer, where the air is simply too dry and stable in THEO, such that they dilute the unstable parcels coming from within the boundary layer. This is in keeping with the lower magnitude of MLCAPE and higher magnitude of MLCIN in THEO. Additionally, the greater free-tropospheric moisture in MODTHEO may result in less dilution of updraft parcels throughout their ascent such that they realize more of their CAPE, which is consistent with the findings of \citet{James2010}.

Ultimately, though, our experiments are \textit{not} intended to demarcate robust sensitivities, nor can we can we cleanly attribute differences in qualitative behavior to any specific feature of the sounding (e.g. boundary layer vs. lower-tropospheric vs. mid-tropospheric moisture) or to changes in specific bulk parameters such as MLCAPE. Instead, our results motivate how the model could be used as the basis for comprehensive testing of the role of these detailed variations in vertical thermodynamic structure in the SCS outcome. Such an approach would require in-depth experimentation via experimental ensembles and consideration of a range of carefully-defined soundings, whether semi-theoretical (akin to MODHIST) or fully-theoretical (akin to MODTHEO). This effort lies beyond the scope of this work. Here we focus simply on presenting the model construction and demonstrating how it could be used to improve our understanding of the effects of any type of variability in a sounding in a simplified setting.


\section{Conclusions}\label{sec:conclusion}

Severe convective storm activity depends not only on bulk parameters such as CAPE and lower-tropospheric shear but also on the detailed vertical structure of the thermodynamic and kinematic profiles that can vary independently of those bulk parameters. Past simulation work has tested these dependencies using the Weisman and Klemp idealized thermodynamic profile model, whose simple parametric construction was motivated by practical utility for SCS research. A preferable alternative would be a model whose structure is defined on physical grounds, and in a manner consistent with how such environments are generated within the climate system. Such a model could be a useful tool for understanding how SCS evolution depends on the vertical structure of the hydrostatic background environment (i.e. towards smaller scales), as well as how SCS environments depend on the process-level energetics of the hydrostatic atmosphere (i.e. towards larger scales).

Here we have presented a simple physical model for the combined steady thermodynamic and kinematic profiles associated with severe convective storm environments. The thermodynamic component of the model builds off of the two-layer static energy framework proposed by \citet{Agard_Emanuel_2017}. The model superposes a boundary layer with constant moist and dry static energy and constant southerly shear beneath a free tropospheric layer with dry static energy increasing linearly with height (allowing a sub-dry-adiabatic lapse rate), constant relative humidity, and pure westerly shear. A step-function increase in dry static energy, which represents a capping inversion that scales with convective inhibition, is imposed across the boundary layer top. The model is topped off with a dry isothermal stratosphere that defines the tropopause temperature. Overall, the thermodynamic and kinematic components are mutually consistent, as they represent a transition from predominantly southerly flow advecting warm, moist (i.e. high moist static energy) air near the surface to predominantly westerly flow advecting warmer, dry (i.e. high dry static energy) air aloft. This static energy framework provides greater physical insight into the ad-hoc structure of the Weisman and Klemp sounding while offering novel benefits, particularly the explicit representation of a capping inversion at the interface between the boundary layer and the free troposphere, each of which may be varied independently.

To demonstrate its experimental utility, we then provided an algorithm for creating a model sounding as well as for fitting the model to a real-data sounding associated with the 3 May 1999 Oklahoma tornado outbreak. Using numerical simulation experiments, the real-data sounding produces a long-lived supercell, whereas our theoretical sounding produces only a short-lived storm. We then demonstrate two specific types of experiments with our theoretical model that also simulate a long-lived supercell: 1) experiments using semi-theoretical soundings that test the importance of specific features in real soundings by incorporating them directly into the model (here we matched the low-level moisture); and 2) experiments using fully-theoretical soundings that test direct variations in the model's physical parameters (here we enhanced the free-tropospheric relative humidity). These two types of experiments demonstrate the potential utility of our theoretical model for testing how SCS evolution depends on details of the vertical structure of a sounding.

This work has focused narrowly on presenting for the community the motivation for the model, how the model is constructed, and how it can be applied to a real-data sounding for sensitivity testing and controlled experimentation. The specific outcomes of our simulation examples shown here are \textit{not} intended to demonstrate robust sensitivities. Such tests of any individual parameter or structural feature requires careful experimentation accounting for key sensitivities in model design via ensemble simulations to ensure real, systematic variability and to falsify alternative hypotheses.

Importantly, the specific construction of our theoretical model as presented here should not be interpreted as final. Instead, this framework should be viewed as a flexible minimal model sufficient to define a viable SCS sounding  -- i.e. one with substantial CAPE and vertical wind shear, and relatively low CIN. Note that this does not guarantee that any specific SCS type (e.g. supercell) will form for a given set of model parameters. This is a natural base model that can be used to test hypotheses regarding SCS environmental dependencies. Experiments could vary vertical thermodynamic structure at fixed CAPE or vertical kinematic structure at fixed bulk shear over different shear layer depths. Structural features may be added or modified as needed; there is no single ``correct'' model.  We hope that future research testing the model and modifications to it may identify other features that are essential to SCS morphology and evolution and thus may be incorporated into this minimal model for practical applications to understanding the diverse range of SCS types on Earth. Moreover, we note that this modeling framework may potentially be adaptable to other types of convective scenarios, such as nocturnal convection.

Finally, the phrasing of the model in terms of moist static energy aligns neatly with the field of climate physics, whose principal focus is the global and regional energy budget of our hydrostatic atmosphere. Hence, the model may provide a useful tool for understanding how and why SCS environments are produced within the climate system in the first place. Understanding how SCS activity will change in a future climate state depends on understanding not only changes in bulk parameters such as CAPE but also changes in the vertical thermodynamic and kinematic structure within favorable SCS environments.


\vspace{10mm}
\noindent\textit{Data Availability Statement.} All data and code from this manuscript are available by emailing the corresponding author at drchavas@gmail.com.

\vspace{5mm}

\acknowledgments
The authors thank John Peters, Matt Gilmore, and one anonymous reviewer for their useful feedback that helped improve this manuscript. This work also benefited from discussions with Tim Cronin and Mike Baldwin. Chavas was partially supported by NSF grant 1648681 and NOAA grant NA16OAR4590208. Dawson was partially supported by NOAA grants NA16OAR4590208 and NA18OAR4590313. The authors gratefully acknowledge the open-source Python community, and particularly the authors and contributors to the Matplotlib \citep{Hunter2007} and MetPy \citep{metpy} packages that were used to generate many of the figures. Computational resources were provided by the Purdue RCAC Community Cluster program.



 \appendix[A] 

\appendixtitle{Potential temperature expression for an isothermal layer}

Here we show how the potential temperature equation above the tropopause in the WK sounding (second equation in Eq. \eqref{eq:theta_WK}) yields an isothermal layer. This result is obtained by first taking the natural logarithm of the definition of potential temperature (Eq. \eqref{eq:pottemp}) to yield
\begin{equation}
ln\theta = ln T - \frac{R_d}{C_p}(ln P - ln P_0)
\end{equation}
Taking a differential yields
\begin{equation}
dln\theta = dln T - \frac{R_d}{C_p}dln P
\end{equation}
We may use the Ideal Gas Law and hydrostatic balance (Eq. \eqref{eq:DALR}) to rewrite the log-pressure differential term $dln P = -\frac{g}{R_dT}dz$. Substituting in and rearranging yields
\begin{equation}
dln\left(\frac{\theta}{T}\right) = \frac{g}{C_pT}dz
\end{equation}
Integrating both sides from the tropopause upwards yields
\begin{equation}
ln\left(\frac{\theta}{T}\right) - ln\left(\frac{\theta_{tpp}}{T_{tpp}}\right) = \frac{g}{C_p}\int_{z_{tpp}}^z\frac{1}{T}dz'
\end{equation}
which can be rewritten as
\begin{equation}
\theta = \theta_{tpp}\left(\frac{T}{T_{tpp}}\right) exp\left[\frac{g}{C_p}\int_{z_{tpp}}^z\frac{1}{T}dz'\right]
\end{equation}
Taking $T(z)=T_{tpp}$ constant (i.e. isothermal) yields
\begin{equation}
\theta = \theta_{tpp} exp\left[\frac{g}{C_pT_{tpp}}(z-z_{tpp})\right]
\end{equation}
which matches Eq. \eqref{eq:theta_WK} above the tropopause.


 \appendix[B] 

\appendixtitle{Saturation fraction vs. relative humidity}

For Earth-like atmospheres in which the saturation vapor pressure is small compared to the total pressure, i.e. $e^* \ll P$, one can show that the saturation fraction and relative humidity of an air parcel are nearly identical. As a result, the two quantities will also be nearly identical for any mass-weighted layer average.

Relative humidity is defined as
\begin{equation}
    RH = \frac{e}{e^*}
\end{equation}
Saturation fraction is defined as
\begin{equation}
    SF = \frac{q}{q^*}
\end{equation}
These equations may be combined with the relations
\begin{equation}
    q = \frac{r}{1+r}
\end{equation}
and
\begin{equation}
    r = \epsilon\frac{e}{P-e}
\end{equation}
and their saturated counterparts, where $\epsilon = \frac{R_d}{R_v} = 0.622$ is the ratio of specific gas constants for dry air and water vapor. The result may be written as
\begin{equation}
    SF = RH\left(\frac{1-\frac{(1-\epsilon)e^*}{P}}{1-\frac{(1-\epsilon)e}{P}}\right)
\end{equation}
Thus if $(1-\epsilon)e^* \ll P$, then $SF \approx RH$. This easily holds for the modern Earth atmosphere, for which at very warm temperatures $(1-\epsilon)e^* \approx (1-0.622)(0.5 \; hPa) \approx 0.2 \; hPa$ which is several orders of magnitude smaller than the associated surface pressures of 1000 hPa. Indeed, the vertical profiles of SF and RH are indistinguishable for the 3MAY99 sounding presented here.

\bibliographystyle{ametsoc2014}
\bibliography{refs_CHAVAS,refs_DAWSON}

\end{document}